\documentclass[aps,prb,twocolumn,superscriptaddress, longbibliography]{revtex4-2}

\usepackage{graphicx}
\usepackage{dcolumn}
\usepackage{bm}
\usepackage{hyperref}
\usepackage{todonotes}
\usepackage{color}
\usepackage{glossaries}

\def\lbcox{La$_{2-x}$Ba$_x$CuO$_4$}
\def\lscox{La$_{2-x}$Sr$_x$CuO$_4$}
\def\lbco{La$_{1.875}$Ba$_{0.125}$CuO$_4$}
\def\ybco{YBa$_2$Cu$_3$O$_{6+x}$}
\def\um{\text{µ}\text{m}}

\newacronym{CDW}{CDW}{charge-density wave}
\newacronym{LTT}{LTT}{low-temperature tetragonal}
\newacronym{LTO}{LTO}{low-temperature orthorhombic}
\newacronym{HTT}{HTT}{high-temperature tetragonal}
\newacronym{FWHM}{FWHM}{full-width at half-maximum}
\newacronym{CIF}{CIF}{Crystallographic Information File}
\newacronym{RLU}{RLU}{reciprocal lattice units}
\newacronym{irrep}{irrep}{irreducible representation}

\begin{document}

\title{Structure of Charge Density Waves in La$_{1.875}$Ba$_{0.125}$CuO$_4$}

\author{J. Sears}
\author{Y. Shen}
\affiliation{Department of Condensed Matter Physics and Materials Science, Brookhaven National Laboratory, Upton, New York 11973, USA}

\author{M. J. Krogstad}
\affiliation{Materials Science Division, Argonne National Laboratory, Lemont, IL 60439, USA}
\affiliation{Advanced Photon Source, Argonne National Laboratory,
Argonne, Illinois 60439, USA}

\author{H. Miao}
\affiliation{Department of Condensed Matter Physics and Materials Science, Brookhaven National Laboratory, Upton, New York 11973, USA}
\affiliation{Material Science and Technology Division, Oak Ridge National Laboratory, Oak Ridge, Tennessee 37830, USA}

\author{E. S. Bozin}
\affiliation{Department of Condensed Matter Physics and Materials Science, Brookhaven National Laboratory, Upton, New York 11973, USA}
\author{I. K. Robinson}
\affiliation{Department of Condensed Matter Physics and Materials Science, Brookhaven National Laboratory, Upton, New York 11973, USA}
\affiliation{London Centre for Nanotechnology, University College,
Gower St., London WC1E 6BT, UK}
\author{G. D. Gu}
\affiliation{Department of Condensed Matter Physics and Materials Science, Brookhaven National Laboratory, Upton, New York 11973, USA}

\author{R. Osborn}
\author{S. Rosenkranz}
\affiliation{Materials Science Division, Argonne National Laboratory, Lemont, IL 60439, USA}

\author{J. M. Tranquada}
\author{M. P. M. Dean}
\email{mdean@bnl.gov}
\affiliation{Department of Condensed Matter Physics and Materials Science, Brookhaven National Laboratory, Upton, New York 11973, USA}

\date{\today}

\begin{abstract}

Although \gls*{CDW} correlations exist in several families of cuprate superconductors, they exhibit  substantial variation in \gls*{CDW} wavevector and correlation length, indicating a key role for \gls*{CDW}-lattice interactions. We investigated this interaction in \lbco{} using single crystal x-ray diffraction to collect a large number of \gls*{CDW} peak intensities, and determined the Cu and La/Ba atomic distortions induced by the formation of \gls*{CDW} order. Within the CuO$_2$ planes, the distortions involve a periodic modulation of the Cu-Cu spacing along the direction of the ordering wave vector. The charge ordering within the copper-oxygen layer induces an out-of-plane breathing modulation of the surrounding lanthanum layers, which leads to a related distortion on the adjacent copper-oxygen layer. Our result implies that the \gls*{CDW}-related structural distortions do not remain confined to a single layer but rather propagate an appreciable distance through the crystal. This leads to overlapping structural modulations, in which CuO$_2$ planes exhibit distortions arising from the orthogonal \glspl*{CDW} in adjacent layers as well as distortions from the \gls*{CDW} within the layer itself. We attribute this striking effect to the weak $c$-axis charge screening in cuprates and suggest this effect could help couple the \gls*{CDW} between adjacent planes in the crystal.
\end{abstract}

\maketitle

\section{Introduction}

Although the nature of superconductivity in cuprates remains controversial, there is increasing evidence that charge and spin correlations might play an important role \cite{kive03, fuji12a, frad15, keim15}. Experimental studies have shown that \gls*{CDW} correlations are ubiquitous in underdoped cuprates \cite{Hucker2011stripe,fink11,gupt21,ghir12,huck14,dasi14,Thampy2014rotated,comi14,tabi17,kohs07,webb19}, while modern numerical calculations suggest that \glspl*{CDW} are an intrinsic feature of minimal model Hamiltonians relevant to these materials \cite{Zheng2017stripe, Huang2017numerical,huan18, jian21c}. Cuprates such as \lbcox{} are composed of two-dimensional CuO$_2$ layers which support \gls*{CDW} order, separated by electronically inert buffer layers. While the \gls*{CDW} is primarily an electronic phenomenon, the rearrangements of charge associated with the \gls*{CDW} necessarily involves atomic displacements, which can play a substantial role in stabilizing the order. For example, measurements of phonons show a strong coupling between \glspl*{CDW} and the in-plane Cu-O bond stretching phonon mode \cite{pint04,rezn06,tejs20,peng20,Lin2020strongly,  wang21a,lee21a}. Furthermore in \lbcox{}, the material of interest for this study, static \gls*{CDW} order only develops below the \gls*{LTT} structural phase transition \cite{fuji04,Hucker2011stripe,wilk11,Miao2017high, Miao2018incommensurate, Miao2019formation}.   Within this phase, the O atoms along one of the Cu-O bond directions are displaced out of the plane (in alternating directions), while in the orthogonal direction the O atoms remain within the CuO$_2$ planes. This broken rotational symmetry has long been believed to pin \gls*{CDW} order \cite{tran95a}, leading to a short-range \gls*{CDW}-structure interaction that rotates $90^\circ{}$ from plane-to-plane due to the screw axis symmetry. \lscox{}, which lacks the LTT structural phase, supports a \gls*{CDW} order that is far weaker than that of \lbcox{} \cite{Thampy2014rotated, crof14}. Further evidence of \gls*{CDW}-structure coupling is seen in the temperature-dependence of \gls*{CDW}-domain formation, which demonstrates that the CDW domains are pinned by the structural domain walls \cite{Chen2019}. In this work, we use x-ray scattering to determine the atomic displacements associated with the \gls*{CDW}.

Advances in synchrotron x-ray instrumentation have made it practical to measure superlattice peaks over a large volume of reciprocal space \cite{krog20}. In the present paper, we exploit these developments to collect an extensive set of diffraction data from the \gls*{CDW} Bragg peaks in \lbco{}.  From a fitting analysis, we are able to determine the \gls*{CDW}-induced displacements of the Cu and La atoms. Despite the participation of O atoms in the \gls*{CDW}, we do not comment on O displacements, due to the low sensitivity of x-rays to O atoms. We find that charge order induces a longitudinal modulation of Cu-Cu spacings within the CuO$_2$ plane. We also find that the La atoms show substantial displacement along the $c$ axis, contributing to screening of the charge order. Remarkably, the modulation of the La layers leads to a similar modulation in the nearest-neighbor CuO$_2$ layer (transverse to the direction of its \gls*{CDW} wave vector). We attribute these long-range \gls*{CDW} effects to the weak $c$-axis charge screening in cuprates. The propagation of the distortions through the crystal helps to communicate the stripe orientation between the next-nearest-neighbor layers.  Lastly, we find evidence that the \gls*{CDW} modulation is aligned along the direction of the tilted in-plane Cu-O bonds.

\section{\label{sec:data}Data Collection}
La$_{1.875}$Ba$_{0.125}$CuO$_4$ single crystal samples were grown at Brookhaven National Laboratory using the floating zone method. These crystals are from the same source that was used previously \cite{Hucker2011stripe} and exhibited a consistent \gls*{CDW} transition temperature. The samples were cut with a razor blade and a piece approximately 200~\um{} in all directions was selected for the measurement. X-ray diffraction data were collected at the 6-ID-D beamline at the Advanced Photon Source at the Argonne National Laboratory. The beam size was $500\times500$~\um{}$^2$ and the x-ray energy was set to 87~keV to collect a large volume of reciprocal space and to minimize signal distortions due to absorption. For \lbco{} the attenuation length for 87~keV x-rays is $\sim600$~\um{}. The samples were chosen to be smaller than the beam size and isotropic in shape in order to mitigate variation in intensity due to absorption effects or the amount of sample in the beam. We selected a sample with a mosaic spread of less than the $0.1^\circ$ experimental resolution for the data collection. 

The sample was cooled using a helium cryocooler, and data were collected by continuously rotating the sample while reading out the Pilatus 2M CdTe detector every $0.1^{\circ}$. The images were processed by mapping each pixel into reciprocal space and then binning the pixels into a 3-dimensional grid of voxels. Bright Bragg peaks were masked on the raw images prior to binning to improve the visibility of weak features such as the \gls*{CDW} peaks. Two further different rotations were made at different sample tilt angles, in order to provide comprehensive mapping of reciprocal space and to fill in unpopulated voxels. The data shown correspond to a total measurement time of 4 hours. 

\begin{figure}[!hbtp]
\includegraphics[width=.95\columnwidth]{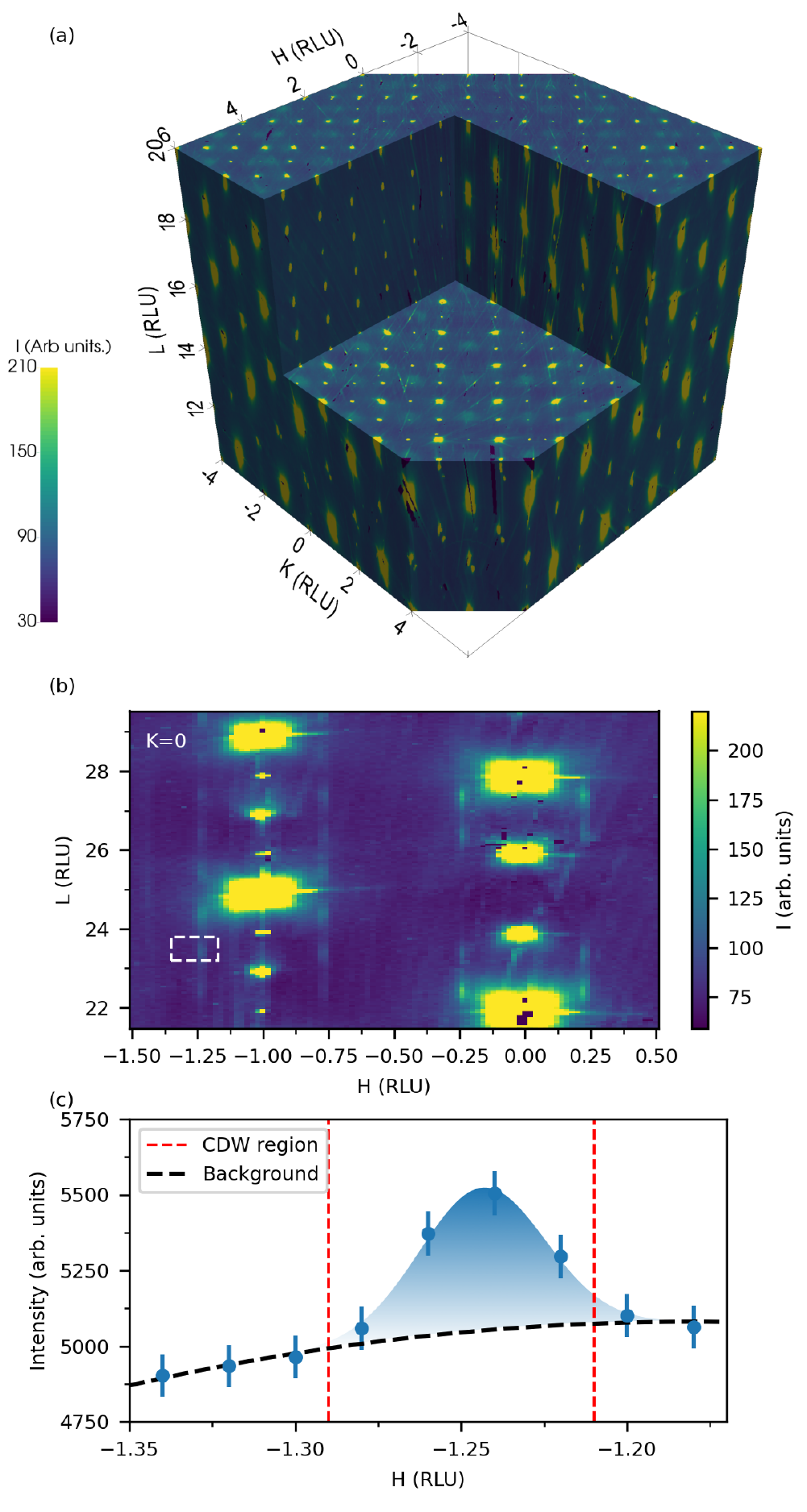}
\caption{\lbco{} \gls*{CDW} measurements over a large reciprocal space range at 28~K. {(a) Extended reciprocal space data ranging from $L=10$ to $L=20$, showing diffuse scattering from the \acrfull*{HTT} structure (at integer positions) and \acrfull*{LTT} peaks at ($\frac{1}{2}$, $\frac{1}{2}$, $L$) positions. (b) $(H0L)$ reciprocal space map showing Bragg peaks and \gls*{CDW} signal appearing as diffuse rods located at $\pm 0.24$ to either side of the Bragg peaks. (c) The region indicated by the white rectangle in panel (b) summed over the $K$ and $L$ directions. The background was fit with a quadratic function over a small region surrounding the known peak position. Four voxels covering 0.08 \acrfull*{RLU} were summed over to encompass the \gls*{CDW} peak width, as illustrated by the vertical red dashed lines.  The points within this region were  summed to give the \gls*{CDW} peak intensity. Error bars shown are the square root of the total photon counts. The faint streaks that form along random reciprocal space directions are detector artifacts as described in the text.}}
\label{recmap}
\end{figure}

Figure~\ref{recmap}(a) illustrates the large extent of reciprocal space measured, and shows approximately one-sixth of the full dataset. The data were collected at 28~K, well below the simultaneous \gls*{LTT} structural phase transition and \gls*{CDW} ordering at 54~K. Throughout this work we will refer to reciprocal space positions using \acrfull*{RLU}, in terms of the unit cell of the \gls*{HTT} phase with $a=b=3.78$~\AA{} and $c=13.19$~\AA{}. We use this notation to maintain consistency with literature, even though \lbco{} goes through two structural transitions with decreasing temperature, entering the \gls*{LTO} phase at 240~K and the  \acrfull*{LTT} phase at 54~K. The unit cell of the  \gls*{LTO} and  \gls*{LTT} phases is larger and rotated by $\sim 45^{\circ}$, leading to superlattice peaks at $(\frac{1}{2}, \frac{1}{2}, L)$ positions. These superlattice peaks are visible in the horizontal cuts of the reciprocal space shown in Fig.~\ref{recmap}(a).

\begin{figure*}[!hbtp]
\includegraphics[width=2.\columnwidth]{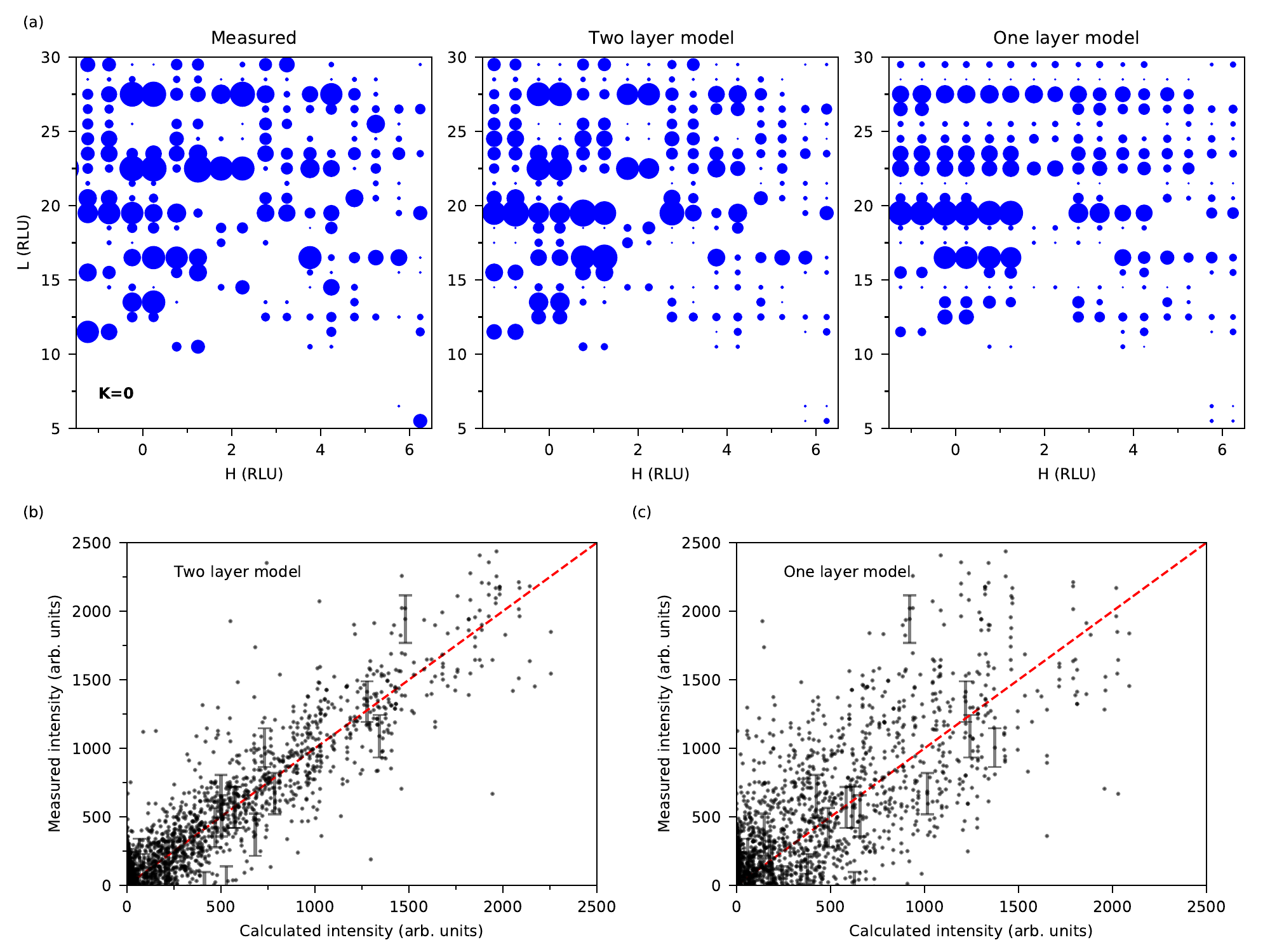}
\caption{
Comparison between measured \gls*{CDW} diffraction intensity and structural models with modulation of either one or both formula unit layers of the unit cell. (a) Plots of the $K=0$ diffraction planes. The left panel shows the experimental data.  Blank areas in the diffraction planes denote areas where a \gls*{CDW} intensity could not be definitely extracted. The middle and right panels shows the calculated peak intensities for the $Bbmb$ space group of \acrfull*{irrep} S1 using modulations of either one or both planes in the crystal structure (one-layer and two-layer models respectively). In all panels, the radius is proportional to the intensity. (b) and (c) Measured vs.\ calculated intensity for the best fit values of the model parameters. The dashed red line traces the ideal case of the measured and calculated intensity being equal. The better fit in panel (b) proves that the \gls*{CDW} induces distortions in both planes of the \lbco{} unit cell. For readability, the errorbars are shown for only 1\% of the data points.}
\label{fig2}
\end{figure*}

The data also show faint but clearly visible \gls*{CDW} signal at wavevectors of $\vec{q}=\pm (0.24, 0, 0.5)$ and $\vec{q} = \pm (0, 0.24, 0.5)$ surrounding the structural Bragg peaks, as shown in Fig.~\ref{recmap}(b). The \gls*{CDW} peaks are extended and rod-like in the $L$ direction, and are centered at half-integer $L$ positions. These are well-known features of the \gls*{CDW} in \lbco{} --- all detectable \gls*{CDW} peaks are of the type $\vec{q} = (\pm0.24, 0, \pm0.5)$ or $\vec{q} = (0, \pm 0.24, \pm 0.5)$ \cite{Kim2008, wilk11, Hucker2011stripe}. The rod-like structure is due to the short correlation length (a few unit cells) in the out-of-plane direction, while the half-integer $L$ peak position is a result of the out-of-phase stacking of the charge pattern between adjacent unit cells. While the low resolution of our measurement precludes a precise estimate of the \gls*{CDW} correlation lengths, our observations are consistent with the previously reported anisotropic correlation lengths of approximately 200 \AA{} in-plane and a few unit cells out of plane.

The intensities of the \gls*{CDW} peaks were extracted from this data set by summing over each four-voxel-wide  region surrounding the peaks and subtracting the background as shown in Fig.~\ref{recmap}(c). Despite masking of bright Bragg peaks, some streaks of intensity remain in the data due to `blooming' from saturation of the detector. The reciprocal space map was manually inspected in the region of each peak to check for bright streaks at the \gls*{CDW} peak position, and contaminated peaks were excluded from the data set. The error in the peak intensity was estimated as the square root of the intensity. Due to twinning, and lattice symmetry, the \gls*{CDW} produces peaks along both $H$ and $K$, so for simplicity we can choose to focus on either $H$ and $K$ without losing any information. For the analysis, we chose to analyze peaks with the \gls*{CDW} along $H$ i.e.\ peaks with $H \approx n \pm 1/4$ where $n$ is an integer.  The data were also symmetrized by reflection in the $K=0$ mirror plane.

\section{\label{sec:analysis}Structural Model and Fitting}

While there are \gls*{CDW} modulations in two orthogonal in-plane directions, there is no evidence for significant coupling between these modulations.  This is consistent with the \gls*{CDW} orientation being tied to the in-plane structural anisotropy of the \gls*{LTT} structure.  Hence, we will analyze the modulation for just one direction, assuming that equivalent results apply for the second. We emphasize that the full crystal structure will necessarily include distortions due to both CDW components. Since no higher harmonic \gls*{CDW} peaks have ever been observed, the structural models were constructed assuming simple sinusoidal modulations of the atomic positions. 

The software Isodistort \cite{isodist, campbell_2006} was used in our analysis to identify CDW component structures compatible with the different types of symmetry breaking that may occur in the \gls*{LTT} crystal structure. In the analysis, we simulated our observed set of \gls*{CDW} Bragg peaks assuming a commensurate wave vector $\vec{q}=(0.25,0,0.5)$ from the \gls*{CDW} domains that lie along $H$, and identified four possible \glspl*{irrep} of the parent $P4_2/ncm$ space group labeled S1, S2, S3, and S4. These \glspl*{irrep} denote the different ways that the distortion can break the crystal symmetry, grouping together distortion modes that maintain a given symmetry operation. S1 and S3 both maintain a vertical mirror plane symmetry which is broken in S2 and S4. We will show later that the data support the preservation of the vertical mirror plane. S1 and S3 differ in the relative orientation of the \gls*{CDW} modulation with respect to the bent in-plane Cu-O bonds within the \gls*{LTT} phase. Structural refinements, which we will also outline later, support the S1 CDW component structure, which has the \gls*{CDW} modulation parallel to the bent in-plane Cu-O bonds. Isodistort files describing the atomic displacements involved in the different symmetry modes are provided in the Supplemental Material \cite{supp}.

The individual distortion modes are sinusoidal modulations of atomic positions along the direction of the ordering vector. Phase transitions often involve only a single irreducible representation, so each of these possible CDW component structures was refined individually against the experimental data. Each of these models includes only a subset of the free parameters required to describe the atomic positions within a suitable supercell, greatly simplifying the fitting procedure. The distortion mode model of Isodistort was converted to the crystal structure, which was used to compute the \gls*{CDW} peak intensities for comparison with the experimental data. The atomic form factors provided by Waasmaier \emph{et al}.\ \cite{waasmaier_1995} were used. The overall scale factor for the data was determined by modeling the intensity of 240 weak Bragg peaks (using the known \gls*{LTT} structure) and comparing with the observed intensity. A commensurate ordering vector was used for the symmetry analysis, despite the fact that the actual wave vector of the scattering is slightly incommensurate. It is thought that this shifted wave vector arises from a mixture of locally commensurate charge stripe structures \cite{tranquada_1999, Miao2019formation}, and we will show later that the measured diffraction patterns are minimally sensitive to the phase of the \gls*{CDW} with respect to the underlying lattice. 

We also note that since the \gls*{CDW} exists within the \gls*{LTT} structure, symmetry dictates that additional \gls*{CDW} peaks of the form $\vec{q}=(0.5\pm 0.24,0.5,0.5)$ should exist. These are indeed present in our model, but with an intensity several orders of magnitude weaker than the $\vec{q}=(0.24,0,0.5)$ peaks, consistent with the absence of observable peaks of this type in the experimental data.

The CDW component structural models of \glspl*{irrep} S2 and S4 were found to predict very little intensity in the $K=0$ plane. Our experimental data show that the brightest peaks lie in this plane, and so these \glspl*{irrep} could be immediately discarded. S1 and S3 each allow a number of possible distorted structures, and the two highest symmetry structures for each \glspl*{irrep} were refined against the experimental data for a total of four candidate structures. For both S1 and S3, the two highest symmetry distortions had space groups $Bbmb$ and $Bmmb$. Although these space groups are both orthorhombic, we note that this is a consequence of including only one of the two \gls*{CDW} domains, and does not reflect the symmetry of the full crystal structure (which will include both \gls*{CDW} ordering vectors). We do not observe any difference between the $a$ and $b$ lattice constants within the experimental resolution, however, we note that our measurement is not sensitive to small changes in lattice constants.

\begin{figure*}[!hbtp]
\includegraphics[width=1.9\columnwidth]{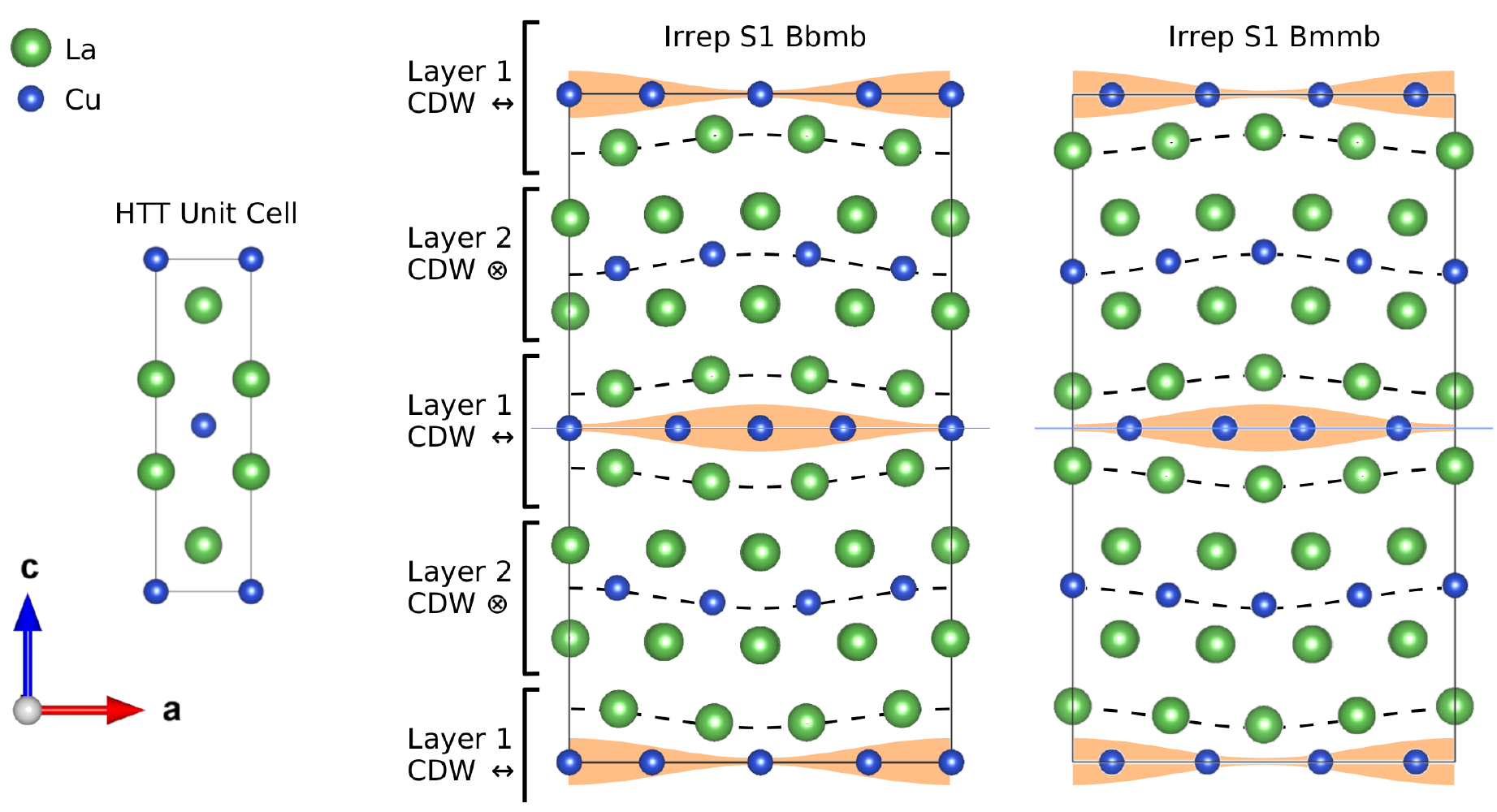}
\caption{
The two refined CDW component structural distortions found to best fit the data corresponding to the $Bbmb$ and $Bmmb$ space groups of the S1 \glspl*{irrep}. These structures show the modulation of the atomic positions away from the flat planes of the \gls*{HTT} structure due to the \gls*{CDW} propagating horizontally in Layer 1, as denoted by $\leftrightarrow$. Layer 2 also hosts a \gls*{CDW} propagating into the page. For clarity this is not shown, but it is denoted by $\otimes$.  The modulations result in a $4\times2\times2$ supercell compared with the \gls*{HTT} structure, as shown by the black outlines. The \gls*{CDW} order causes in-plane modulation of the copper positions on the central plane (blue solid line), and a breathing type distortion is transmitted to the surrounding layers of atoms (modulation shown by dashed lines). Remarkably, the distortion involves not only Layer 1, which hosts the \gls*{CDW}, but also the adjacent Layer 2. Based on the observed distortion, the probable pattern of hole density modulation is illustrated by the orange shading. The two refined distortions are very similar, differing only in the phase of the modulation with respect to the crystal structure. The size of the distortion has been increased by a factor of 100 for visibility, and the structures were plotted using Vesta \cite{vesta}.}
\label{fig3}
\end{figure*}

Initial modelling using each of the candidate CDW component structures was done allowing all of the La and Cu $x$ and $z$ modulation amplitudes to vary. Along the reciprocal space $K$ direction the peak intensities decrease rapidly, suggesting that the distortions in the $y$ direction perpendicular to the \gls*{CDW} are small compared to displacements along $x$ and $z$. Attempts to refine the $y$ displacements did not lead to statistically significant improvements to the fit so these modulations were therefore fixed to zero. Several of the refined modulations were found to be small in comparison to their computed errors. These values were also fixed to zero, resulting in minimal models with only 6 free parameters for each of the candidate structures. Adding further parameters to these models did not substantially improve the goodness of fit. In particular, we performed fits in which the La/Ba atoms were fixed and the Cu and O were free, but these failed to reproduce the data. Fits in which all atoms (La/Ba, Cu, and O) were free resulted in only minor improvements in the fit considering the large number of additional parameters added. This is not surprising considering that the ratio of atomic numbers of La to O is 57:8, so La contributes $(57/8)^2\approx 51$ times more strongly to the x-ray scattering for a similar displacement. The oxygen distortions were therefore not included in the final refinement.

\section{\label{sec:discussion}Results and Discussion}

The unit cell of \lbcox{} contains two \lbco{} formula units, related at low temperature by a screw axis parallel to the crystal $c$ axis. Each formula unit layer consists of a CuO$_2$ plane and its surrounding lanthanum atoms. Throughout this discussion, we will refer to layers with the \gls*{CDW} ordering wave vector along the direction of the bent Cu-O-Cu bonds as `Layer 1'. The remaining formula unit layers, with the wave vector along the straight Cu-O-Cu bonds, will be referred to as `Layer 2'. Fig.~\ref{fig3} shows how the crystal structure is constructed out of alternating layers 1 and 2. These two layer types are thought to account for the two \gls*{CDW} domains, with the ordering vector in adjacent layers rotated by 90$^{\circ}$, consistent with the screw axis symmetry. The data chosen consists of peaks from only one of the two domains, and so we initially considered structural models in which the \gls*{CDW} induced distortion was confined to a single layer type. Regardless of the structural model used, allowing only a single layer to distort produced only very poor fits to the data, and failed to reproduce the qualitative features seen in the diffraction pattern as shown in Fig.~\ref{fig2}. 

Good fits to the data could only be achieved by allowing both layers to distort. When this was done, the qualitative features of the diffraction pattern could be reproduced with all four of the remaining CDW component structures: space groups $Bbmb$ and $Bmmb$ for each of the S1 and S3 \glspl*{irrep}. In all cases the $\chi^2$ values were $\sim 1$, however the fits for the S3 structures were slightly worse ($\chi^2=1.06$ for both $Bbmb$ and $Bmmb$) in comparison with S1 ($\chi^2=1.02$ for both $Bbmb$ and $Bmmb$). We note that the S1 and S3 CDW component structures all have in common a vertical mirror plane including the ordering vector and the $c$ axis ($ac$ plane in the \gls*{HTT} notation). This mirror plane is also compatible with a truly incommensurate structure, as it leaves the ordering wave vector fixed. In contrast, a symmetry operation (such as a $bc$ mirror plane) which transforms the ordering vector $q$ to $-q$ cannot be simply translated into the incommensurate case.

The refined structures for S1 and S3 are very similar, the only difference being which copper-oxygen plane hosts the charge order. In the S1 structures Layer 1 hosts charge order, with the \gls*{CDW} ordering vector along the direction of the bent copper-oxygen bonds. In the S3 structures Layer 2 hosts the charge order, and the ordering vector is along the straight copper-oxygen bonds. The reason for the similarity in the fits is that the two layers have only small structural differences. Nevertheless, the slight difference in the fit quality suggests that the charge ordering vector is in the direction of the bent copper-oxygen-copper bonds. The refined structures were very similar, and from this point forward we will focus on the S1 structures. 

The refined distortions are on the order of $10^{-3}$~\AA{}, and are given in Table~\ref{tab:table1}.  These values are based on the assumption that the \gls*{CDW} occupies 100\% of the crystal. If the phase fraction is smaller, then the distortions would be rescaled accordingly. Measurements suggest that static charge order occupies the majority of the sample volume, albeit with significant disorder  \cite{savi05, hunt01, chen16, tham17}. Another source of uncertainty is in the overall scaling of the \gls*{CDW} distortions. Due to the small \gls*{CDW} peak intensities, they all scale as the square of the displacements, so this scale factor cannot be independently fit with just the \gls*{CDW} peaks. We determined this scaling by comparing with structural Bragg peaks, which introduces possible uncertainty from self-absorption and multiple-scattering effects. Our use of a sample that is significantly smaller than the x-ray attenuation length means that this uncertainty is smaller than the uncertainty coming from the domain population. The relative sizes of the different distortion modes are not affected by these sources of uncertainty. 

Figure~\ref{fig2}(a) compares the measured diffraction pattern with the intensities predicted by the higher symmetry (space group $Bbmb$) CDW component structure of the S1 \gls*{irrep}. Plots of the measured intensity against the calculated intensity are provided in Fig.~\ref{fig2}(b), such that a hypothetical perfect fit would have points that fall on the red dotted line. We see in Fig.~\ref{fig2}(b) that the fit quality of the model with distortions in both copper-oxygen layers (left panel) is significantly better than case with distortions in just one layer (right panel). The two-layer model is thus able to fit 3500 peak intensities using only 6 free parameters within a scatter comparable to the experimental errorbar. The very similar structure of the S1 \gls*{irrep} with $Bmmb$, rather than $Bbmb$, space group symmetry gives a similar fit to the data. Examining the refined structures (shown in Fig.~\ref{fig3}) suggests the probable reason for this similarity. The $Bbmb$ and $Bmmb$ symmetry structures of the S1 \gls*{irrep} differ only in the phase of the distortion with respect to the crystal structure. Based on the equal goodness of fit, we conclude that this phase cannot be determined based on our data set.

\begin{table}[b]
\caption{\label{tab:table1}%
Maximum CDW-related atomic displacements, with refined values for the $Bbmb$ and $Bmmb$ CDW component structures of \gls*{irrep} S1. The $4\times2\times2$ \gls*{HTT} supercell lattice parameters are $a=15.12$~\AA{}, $b=7.56$~\AA{}, $c=26.38$~\AA{}. The estimated error values in parenthesis are the square root of the diagonal elements of the estimated covariance matrix, as derived from the output of the Scipy leastsq optimization function. Isodistort output files describing the distortion modes of these structures are provided as Supplemental Material \cite{supp}.
}
\begin{ruledtabular}
\begin{tabular}{lcc}
\textrm{Distortion Mode}&
\textrm{$Bbmb$}&
\textrm{$Bmmb$}\\
& parameters & parameters \\
&
\textrm{(\AA$\times 10^{-3}$)}&
\textrm{(\AA$\times 10^{-3}$)}\\
\colrule
A'\_1(a) (Layer 2, La x) & 0.19(1) & 0.13(1)\\
A'\_2(a) (Layer 2, La x) & - & -\\
A'\_3(a) (Layer 1, La y) & - & -\\
A'\_4(a) (Layer 2, La z) & - & -\\
A'\_5(a) (Layer 2, La z) & 0.68(1) & 0.494(8)\\
A'\_6(a) (Layer 1, La z) & 1.356(6) & 1.843(8)\\
A''(a) (Layer 1, La x) & 0.33(1) & 0.49(2)\\
Bu\_1(a) (Layer 1, Cu x) & 2.5(1) & 1.74(8)\\
Bu\_2(a) (Layer 2, Cu y) & - & -\\
Bu\_3(a) (Layer 1, Cu z)& - & -\\
Bu\_4(a) (Layer 2, Cu z)& 1.42(4) & 1.95(6)\\
\end{tabular}
\end{ruledtabular}
\end{table}

Our modeling shows that the \gls*{CDW} in \lbco{} induces a longitudinal modulation of the copper atoms of one of the copper-oxygen planes (Layer 1), likely due to modulation of the hole population within this layer. The charge modulation also leads to a breathing-type distortion in the layers of lanthanum atoms immediately above and below this plane.  The breathing distortion additionally propagates further into the crystal and affects adjacent layers. This finding is remarkable, since the adjacent copper-oxygen layers are expected to support their own charge density wave modulations, rotated by $90^{\circ}$ about the $c$ axis. The propagation of the distortion into these adjacent layers means that in regions of the crystal where the two \gls*{CDW} domains coexist, the structure is described by two wave vectors rather than the simple 1D modulation usually proposed. If the modulation along the charge-stripe direction couples to the charge, then, given a $4a$ commensurability in two directions at $x=1/8$, it might underly the doping dependence of charge order and superconductivity in La$_{2-x}$Ba$_x$CuO$_4$ \cite{huck11,mood88} and La$_{2-x-y}R_y$Sr$_x$CuO$_4$ with $R=$ Nd \cite{ichi00,craw91} and Eu \cite{fink11,sury05} in which the wavevector is essentially independent of doping for dopings above 1/8.

Our results can be compared with previous x-ray diffraction measurements \cite{Kim2008}, which used a structural model of distortions in a single layer to fit a far smaller data set of 70 \gls*{CDW} peak intensities. The authors found that the most important distortions were the longitudinal modulation of copper atoms due to the charge modulation in the copper-oxygen plane, and the modulation of lanthanum atoms along the $c$ axis in response. Our results show that these modulations are indeed significant, but that the breathing modulation also propagates into the adjacent layers.

It is also instructive to compare the distortion in \lbco{} to the previously solved \gls*{CDW} structural distortion in \ybco{} \cite{forgan_2015}. The first notable difference is that the copper-oxygen chain layers of \ybco{} are completely undistorted; in contrast \lbco{} hosts modulations that propagate through the crystal structure leading to modulation in all layers. The distortion of the copper-oxygen square planes also differs between the two materials. Rather than the longitudinal distortions found in the copper-oxygen planes of \lbco{}, the distortions within the \ybco{} square layers are mostly along the out-of plane $c$ direction. Despite these differences, the overall magnitude of the distortions are comparable in both materials. These results highlight the importance of the structural environment and its interaction with electronic order in the copper-oxygen planes. Of course, the zero-field \gls*{CDW} order in \ybco{} has a very short in-plane coherence length and is symmetric about a Cu-O chain layer.  It would be interesting to compare with the high-field 3D \gls*{CDW} order in \ybco{} \cite{gerb15,chan16a}, but this is more challenging to measure.

Since the \gls*{CDW} in cuprates is driven by interactions within the CuO$_2$ planes, our detection of \gls*{CDW}-induced $c$ axis displacements of Cu and La might be considered counter-intuitive. However, the cuprates are known to have very weak electronic screening of Coulomb interactions along the $c$-axis. This is seen in the giant $c$-axis lattice constant change upon photoexcitation \cite{gedi07,rado08} and the large oscillator strength of several $c$-axis-polarized phonon modes in optical conductivity and Raman measurements \cite{uchi96, home12, suga89, goza03}, including some that involve Cu and La motions \cite{henn97}. Inelastic x-ray scattering measurements of \lbco{} also observe a substantial \gls*{CDW}-induced softening of $c$-axis longitudinal acoustic phonons \cite{Miao2018incommensurate}. Prior neutron pair distribution function measurements also associated the \gls*{CDW} with $c$-axis displacements \cite{Bozin2015reconciliation}. It is further worth noting that each La$^{3+}$ ion sits just above (or below) a square of four in-plane O$^{2-}$ ions, so displacing La atoms involves a large modification in the $c$-axis Coulomb interactions. These facts make our observed long-range displacements quite natural.

\section{Conclusions}
Based on a comprehensive measurement of several thousand \gls*{CDW} peaks, we have determined the atomic displacements associated with CDW order in \lbco{}. The structural distortion involves a periodic modulation of the copper atoms along the direction of the \gls*{CDW} and an out-of-plane breathing modulation of the surrounding lanthanum layers. Despite the widespread belief that \gls*{CDW}-related distortions should rotate by $90^\circ$ from layer-to-layer, this breathing modulation, in fact, propagates further through the crystal and induces a similar distortion on the adjacent copper-oxygen layers. We conclude that the effects of charge ordering cannot be considered to be confined to a single plane, but manifest as a complex longer-range distortion that should be described by two wave vectors rather than the simple 1D modulation usually proposed. We attribute this surprising effect to the weak $c$-axis charge screening in cuprates.

The supporting data for the plots in this article are openly available from the Zenodo database \cite{repo}.

\begin{acknowledgments}
We thank C.\ Mazzoli for discussions and D.\ Robinson for assistance with x-ray scattering experiments. Work at Brookhaven is supported by the Office of Basic Energy Sciences, Materials Sciences and Engineering Division, U.S.\ Department of Energy (DOE) under Contract No.\ DE-SC0012704. Work at Argonne National Laboratory (MJK, RO, SR, single crystal diffuse scattering measurements and data reduction) was supported by the U.S. Department of Energy, Office of Science, Basic Energy Sciences, Materials Sciences and Engineering Division. This research used resources of the Advanced Photon Source, a U.S.\ Department of Energy (DOE) Office of Science User Facility operated for the DOE Office of Science by Argonne National Laboratory under Contract No.~DE-AC02-06CH11357.
\end{acknowledgments}

\bibliography{refs,lno,theory}

\end{document}